# A Case Study of Execution of Untrusted Business Process on Permissioned Blockchain

Vahid Pourheidari
Computer Science Department
University of Saskatchewan
Saskatoon, Canada
vahid.p@usask.ca

Sara Rouhani
Computer Science Department
University of Saskatchewan
Saskatoon, Canada
sara.rouhani@usask.ca

Ralph Deters
Computer Science Department
University of Saskatchewan
Saskatoon, Canada
deters@cs.usask.ca

*Many studies have been done to improve the performance of centrally controlled business processes and enhance the integration between different parties of these collaborations. However, the most serious issues of collaborative business processes remained unsolved in these studies – lack of trust and divided data on various confidential ledgers. Blockchain technology has enormous potential to become a new substantial integration method for untrusted collaborative businesses. Using the governing consensus mechanism, blockchain eliminates the necessity of the trusted third party. It provides a distributed shared ledger which facilitates the job of the process monitoring for the parties. The smart contract, as a crucial tool, is used to define the guaranteed autonomous programs. In addition, the privacy of the data can be ensured by using a permissioned blockchain that handles the access control because in this way, only verifiable participants can have access to the state of the business process and its related information. In this study, the applicability of execution of a real-word untrusted business process on the permissioned blockchain is investigated. Moreover, we determine the advantages of using the permissioned access-controller blockchain as the infrastructure for the collaborative business processes, through implementing the process of Order Processing on the Hyperledger Fabric blockchain platform.*

## I. Introduction

Many studies have investigated the effects of the business processes integration on either firms' operations or business performance [1-5]. Therefore, today we know that any level of integration can increase the business performance. However, almost all proposed integration methods face two prominent issues: lack of shared repository and lack of trust.

The first issue is the lack of high-reliability, trusted, tamper-proof, shared ledger. Despite of novel database technologies, communication methods, and encryption techniques, most of the business parties are still store their data and transactions, like hundred years ago, in separate confidential ledgers. This approach makes it hard to track the business's state and the transactions between its parties. Even worse, the collaborative parties may change stored data on their own databases to deceive others and gain more profits.

The second matter is the lack of trust between business's parties. Assume a collaborative business between different organizations, companies or individuals where some of them might be competitors, there are usually many discussions and conflictions to select the orchestrator. In such an untrusted network, who should role as the central controller hub?

Recently, the blockchain technology attracts researchers' and commercial companies' attention. Blockchain inherent features can potentially offer feasible solutions for the mentioned obstacles of collaborative business processes. Accordingly, recent studies have been done on the execution of business processes on the blockchain [6-9] and using the blockchain for the cross-organizational setting and management [10,11]. According to these works, blockchain can cover the lack of trust and the lack of shared ledger for the collaborations.

Blockchain platforms use different consensus mechanisms [12] to validate the transactions in the untrusted network. Therefore, by performing the business processes on the blockchain, there is no need to the trusted hub or the trusted third party anymore. Different parties can join the untrusted network, which could include a set of commercial partners, without any concern about the possible manipulation or corruption. In addition, blockchain provides a distributed tamper-proof shared ledger that can be used for monitoring the process-flow. While every party has a copy of this unparalleled ledger, they can track the state of the process and all executed transactions precisely. Moreover, the complex business and regulatory rules can be reflected in the blockchain platform using smart contracts. Smart contract [13] is a programmable code stored on the blockchain which can enforce its conditions automatically.

Although blockchain creates a trustworthy network that does not need any trusted third party or trust in any single entity, the privacy still is a vital component [6]. Public or permissionless blockchain platforms do not provide any privacy for the data; everybody can join to the public blockchain network and access to the data with no limitation. In contrast, in permissioned blockchain platforms, only approved participants who gained the permission can join the network. However, even the permissioned blockchain is not enough for covering all business processes scenarios because in the permissioned blockchains every joined participant can still see all the data and transactions within the network. A collaborative business process may include various competitors, such as different manufacturers producing the same commodity. Of course, each of the parties want a guarantee that the information about their contracts, assets, and transactions remain invisible to their rivals. In addition, there are many other cases that the data privacy is highly important in the business processes. Hence, to implement the collaborative business process on the permissioned blockchain, it is necessary to have an access control management.

In this study, we investigate the feasibility of the execution of a real-world business process, namely Order Processing, on the permissioned access-controller blockchain. We use the Hyperledger Fabric [19] infrastructure as the blockchain platform and Hyperledger Composer [20] as the development framework. The procedure of the presented study can be classified into the background and related work, business network implementation and execution, and discussion.

## II. BACKGROUND AND RELATED WORKS

### A. Blockchain Technology

Blockchain is a sequence of blocks which include confirmed transactions records. Each block linked to the previous block by adding the hash value of the previous block to the header of its own block. Blockchain works as a distributed network, there is no centralized authority and it runs directly by involved parties without involving any third party. Blockchain initial use case was tackled for cryptocurrency or digital currency trading, however by exploiting blockchain stunning feature, Smart Contract [13], we can develop more variety of applications. Smart Contract is programmable code that resides on the blockchain. By employing smart contracts, we can define more complex transaction and automate the procedures. In Hyperledger Fabric [19] smart contracts is called *ChainCode* since it represents the same functionality.

### B. Bussiness Processes

A business process is "a collection of activities that takes one or more kinds of input and creates an output that is value to the customer" [14]. In other words, a business process is defined with a set of tasks and activities, relations between different activities, clear inputs and outputs, and specific customer(s). A collaborative business process is a group of relevant business processes among different participants that pursue a particular goal.

A business process usually is modeled as a flowchart. One of the most popular standards that provide graphical representations of business processes is the Business Process Modeling and Notation (BPMN) [15]. A BPMN process is made up of diverse types of elements: objects, sequence flows and message flows. An object itself can be an activity, an event, or a gateway. Figure 1 shows the most common elements of the BPMN standard [16].

### C. Order Processing Model

Figure 2 shows the BPMN flowchart of the Order Processing model. In addition, we assign the C, M, and L letters to the business model's activities, start event and XOR decision gateway to specify which party can access to those elements. These letters stand for customer, manufacturer and logistics respectively. Therefore, Figure 1 not only shows the business elements and their sequences, but, it also implies the needed access control rules as well.

There are three kinds of contributors in this business process model: Customer, Manufacturer, and Logistics.

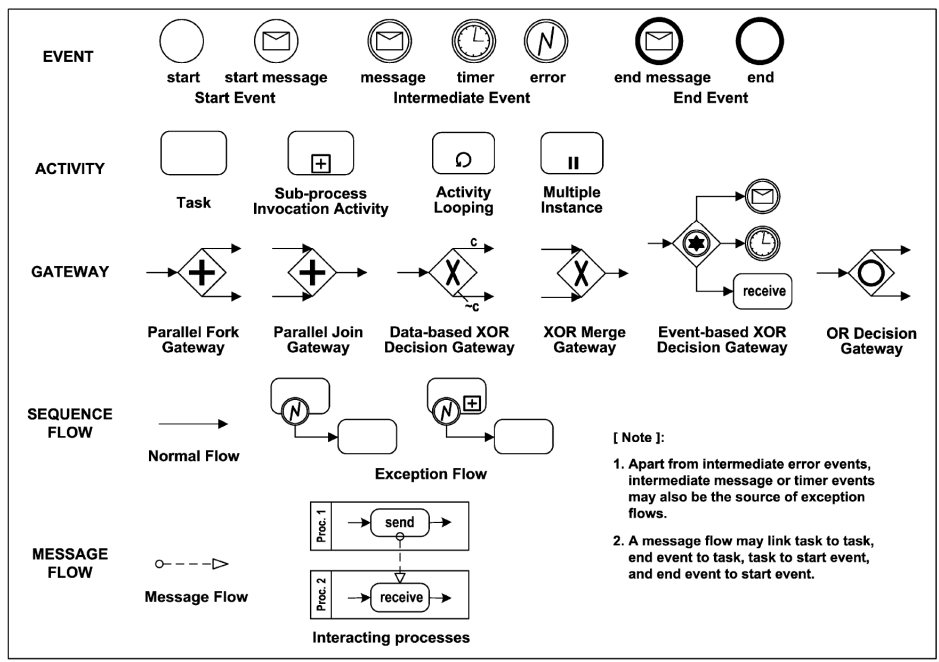

Fig. 1. Elements of BPMN [16].

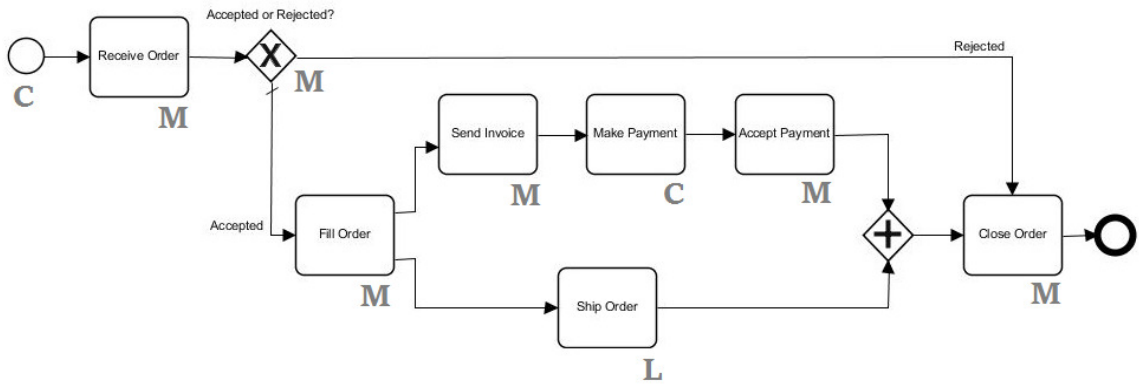

Fig. 2. The Order Processing scenario.

There are three kinds of contributors in this business process model: Customer, Manufacturer, and Logistics. The customer can make an order, to buy a commodity from the manufacturer. The manufacturer can either reject or accept the order after receiving it. If the manufacturer accepts the customer's request, it fills the order and sends the invoice back to the customer. The customer should pay the fees and the manufacturer should check and accept the customer's payment. In addition, after filling the order by manufacturer, the logistics can ship the ordered commodity from manufacturer to the customer. It is a parallel flow, so the Ship Order task can be occurred any time after Fill Order task, independent from Send Invoice, Make Payment and Accept Payment tasks. Finally, if the manufacturer rejects the order or when both the Ship Order and Accept Payment tasks are fulfilled, the manufacturer can close the received order.

### D. Privacy and Access Control in Collaborative Businesses

As it mentioned before privacy is an important part of any practical collaborative business. Hence, we shouldn't consider blockchain as a proper infrastructure for business process management (BPM) if it cannot provide privacy for participants' identities, their assets and contracts, and their transactions. In addition, we need certain access procedures for our deployed smart contracts as well. Consequently, we should be able to define who can have access to one specific resource, transaction, or smart contract in our blockchain network.

Although using the usual centralized access control system seems to simplify the job of the administration, however, there are some shortcoming associated with this method. While we are talking about the centralized approach, the first possible problem that can be implied is the single point of failure. Moreover, in an untrusted collaborative process, allocating the access control authority to one contributor could be problematic and this lack of transparency may constitute further issues.

In this study, the Hyperledger Composer framework is used to define different components of the business network, including access control rules. The Hyperledger Composer comprises an Access Control Language (ACL) that supplies a declarative definition of access control over all resources which are defined in the model file. Using the ACL, we can implement all access control models such as mandatory access control, discretionary access control, role-based access control, and so on. More detail about implementation of business models using the Hyperledger Composer are expressed in the next section.

In addition, the advantages of using decentralized access control system do not limit to the collaborative processes, but, this approach can even use to define the different levels of access in intra-organizational scenarios

### E. Execution of Collaborative Processes on Blockchain

According to the blockchain intrinsic characteristics, which include using consensus algorithms and providing a shared ledger, as well as smart contracts' strength, which equips the blockchain with more complex programmable codes, many studies have been done recently to model blockchain-based business processes.

Weber et al. [6] used the blockchain to address the trust problem in the collaborative business processes. They developed a technique to execute the business processes on both public and private Ethereum [17] blockchains. In their approach, each BPMN model is converted to a factory contract. Then the instances of these factory contracts can be implemented on the blockchain. They used blockchain to either monitoring the process status or coordinating the collaborative process execution. Another feature of their technique is using the triggers and interfaces that connect the off-chain environment to the process execution on the blockchains. Although they correctly show that the blockchain can eliminate the requirement to the trusted third party, but their solution does not completely address the privacy problem of the blockchain networks.

In another study, Garcia-Banuelos et al. [7] suggested an approach to creating minimized Solidity [18] codes from BPMN processes. Solidity is an exclusive language for implementing smart contracts on the Ethereum blockchain. In this method, each BPMN model is first converted to an equivalent Petri Net and after simplification, it is transformed to the Solidity contract code.

Mendling et al. [8], through their paper, enumerated the advantages of the blockchain technology for business process management as well as some likely future directions to union blockchain and BPM. They predicted that the blockchain technology will affect this area significantly.

In addition, many other scientists and researchers have been worked on business process and cross-organizational management systems based on the blockchain technology [9-11]. All these works admit the applicability of using the blockchain to facilitate collaborative process management, but, they also mentioned some concerning conflict points, such as privacy and data access control.

## III. BUSINESS NETWORK IMPLEMENTATION

### A. Defining the Model as a Business Network Archieve

As it said before, the Hyperledger Composer [20] is used as the development framework in this study. This framework is an open development toolset which facilitates the procedures of defining the business network, deploying the created business network, and testing it.

Within the Hyperledger Composer, each business model, for example the order processing model, is defined as a Business Network Archive which includes a set of model files, script files, access control files and optionally query files (Figure 3). In fact, distinct aspects of the business model are defined and stored through these files and then packed as the Business Network Archive which can be deployed on the Hyperledger fabric blockchain platform.

The business domain is defined in the model file by means of the Hyperledger Composer's object-oriented modeling language. The model file comprises the definition of asset(s), participant(s), and transaction(s). For any collaborative business model, we need to define these resources such that the business rules can be followed accurately.

Here, we think about the business flow as an asset including: (i) Boolean properties for each elements of the business model which requires business parties' actions, such as tasks and decision gateways, (ii) relationships to corresponding participants, (iii) and needed properties for describe the content of the business. Therefore, we easily can monitor the business flow and define who can have access to the asset's status and its detail.

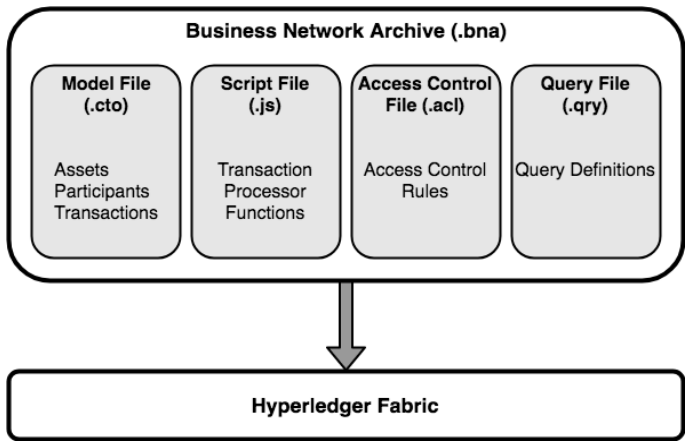

Fig. 3. Elements of a Business Network Archieve

TABLE I. PROPERTIES OF THE ASSET OF THE ORDER PROCESSING MODEL

| Property | Type |
| --- | --- |
| status | Enumerated |
| ID | String |
| name | String |
| description | ID |
| recieveOrder | Boolean |
| rejected | Boolean |
| accepted | Boolean |
| fillOrder | Boolean |
| sendInvoice | Boolean |
| makePayment | Boolean |
| acceptPayment | Boolean |
| shipOrder | Boolean |
| closeOrder | Boolean |
| shopper | Relationship |
| seller | Relationship |
| delivery | Relationship |
| responsible | Relationship |

Table I describes the components of the considered asset for implementing the order processing model.

Status is an enumerated type that can be either active or closed. Once the Customer open an order the status is active; and when the order is closed by the manufacturer, the status is become closed. ID, name and description properties are used to characterize an order. Although they are completely arbitrary, but they are common characteristics that might be selected to describe and recognize different orders. It should be noted that in the Hyperledger Composer, you need an identifying field to define an asset using modeling language. On the other hand, other Boolean and relationship properties are required in our method. The Boolean properties are used to track the process situation. In general terms, the properties with relationship types are unidirectional references to the participant(s). Here, the shopper, seller and delivery relationships are used to specify the customer, manufacturer and logistics of any business network instance, and define the verifiable individuals or legal entities that have access to read a particular asset. In addition, the responsible property, the relationship to the partner(s) which is needed to take an action, is utilized to define which one of the related parties can update the asset in each situation.

Different parties of a collaborative business process can be defined as separate participants in the model file, or ideally, we can define fewer participants in the code and specify their kinds using some enumerated types. By applying this technique, we only need to define one participant in the model file of the order processing model. This participant has a property with enumerated type which can be any of the shopper, seller, or delivery options. Moreover, the participant definition includes some other properties to add the identification characteristics such as first name, last name, ID, company name, and intra-organization position. Like the asset resource, any participant resource need to has an identifying field; and its instances are distinguished by their unique assigned value for this field. Here, we selected the participant's ID property as the identifying field but again all the identification properties are arbitrary and may change based on the content and requirements of the business processes.

In addition to the asset(s) and participant(s), the transactions must be defined in the model file as well. Generally, each transaction definition involves its name, relationships to the assets and/or participants that it needs to work with, and required properties holding new input values. Considering the asset definition of the collaborative business process, we should define one transaction for each Boolean property in order to update its value. In other words, we need transactions for those elements of the BPMN process-flow that needs parties' actions. In addition, we need a unique transaction to initiate new instances of the asset as well. Therefore here, we define nine transactions for their corresponding tasks and XOR gateway of the order processing model and one transaction to create new asset instance.

The script file (Figure 3) contains the implementation of the transactions that are defined in the model file. Each transaction is described by a Transaction Processor Function which includes decorators, metadata, and a JavaScript function. For a collaborative business process, we reflect the sequence of the objects of the BPMN model in these transaction processor functions such that each transaction can be performed if and only if all its required predecessors already were accomplished. For instance, in the order processing model (Figure 2), the Fill Order task can be done if and only if the received order formerly was accepted. Also, the Close Order task can be performed just if the received order was rejected or both Accept Payment and Ship Order tasks were carried out successfully. In addition, each task cannot be performed more than once. Although, the script file explains the detail of transactions and their logical order, but it does not answer to the question that who can perform a specific transaction? This question is answered in the access control file.

Within the access control file (Figure 3), we can define proper access control rules for any collaborative business processes using the ACL. Generally, these rules may contain CREATE, READ, UPDATE, and DELETE operations over all resources which defined in the model file, for any participants of the business network or their instances. The conditional ACL rules, which include Boolean JavaScript expressions, can be used to implement the business norms accurately. Moreover, based on the content and requirement of the business model, different access control methods such as mandatory access control, discretionary access control, or role-based access control may be applied.

For the order processing business model (Figure 2), we establish appropriate access control rules to ensure following principles:

- A new instance of the order processing asset, which is in fact a request for a commodity, only can be created by a shopper (or customer).
- The created instance of the asset can only be seen by its related shopper, seller, and delivery which are different parties of the collaborative business.
- The only one who can perform the corresponding transactions to the Receive Order, Accepted/Rejected, Fill Order, Send Invoice, Accept Payment, and Close Order objects is the related seller (or manufacturer) of the asset instance.
- The related shopper (or customer) of an asset instance is the only one who can execute the corresponding transaction to fulfill the Make Payment task.
- The corresponding transaction to the Ship Order task can only be executed by the related delivery (or logistics) of the asset instance.

Here, we assume that the shopper selects the desired delivery from all different options while creating a new instance of the asset, but it could be defined differently to cover other scenarios as well. For example, the seller might be interested to work with some specific deliveries. Hence, still some implementation detail may vary from case to case.

*B. Deployment and Execution of the Business Network*

As it said before, the business network files are packed as the business network archive and then are deployed on the Hyperledger Fabric blockchain. To determine the validity of the implementation approach and also the functionality of the deployed business network on the blockchain, we set up an experiment.

In this test, we initiated 28 participant instances including 20 shoppers, 5 sellers, and 3 deliveries. Then, 200 asset instances (like Table I) were created randomly over all shoppers, sellers, and deliveries. Considering the asset instances as separate business processes, we defined all valid and invalid transactions for each step of the process flow. The invalid transactions include both intra-process and inter-processes transactions which do not follow the order processing model's rules. In contrast, the valid transactions are those which conform the represented BPMN model's logic from the start event to the end event (Figure 2).

For each asset instances, the test went through valid transactions to complete the business process and fulfill the Close Order task. The answer of the XOR gateway decision in the order processing's BPMN flowchart (Figure 2) was selected randomly, so each asset instance may include either true "rejected" or true "accepted" property (Table I) at the end.

After each valid transaction, the test attempted an invalid transaction. This invalid transaction, could be randomly intra-process or inter-processes transaction.

Finally, we determined the accuracy of our deployed business network in terms of the number of successful valid transaction and also the number of failed invalid transactions. The statistics result of the test are presented in Table II.

TABLE II.  BUSINESS NETWORK TEST'S CONFIGURATION AND RESULT

| | |
|---|---|
| Number of participant instances with shopper type | 20 |
| Number of participant instances with seller type | 5 |
| Number of participant instances with delivery type | 3 |
| Number of all asset instances | 200 |
| Number of asset instances with true "accepted" property | 126 |
| Number of asset instances with true "rejected" property | 74 |
| Number of all valid transactions | 1430 |
| Percentage of successful valid transaction | 100% |
| Number of all intra-process invalid transactions | 817 |
| Number of all inter-processes invalid transactions | 613 |
| Percentage of failed invalid transactions | 100% |

## IV. DISCUSSION AND CONCLUSION

In this research study, we proposed a new method to implement the collaborative business processes, represented by the BPMN standard, on the permissioned access-controller blockchain. Considering the Hyperledger Composer as the development framework and the Hyperledger Fabric as the blockchain platform, our method includes following procedures:

- Defining an asset for the business process flow comprising Boolean properties for each element of the BPMN flowchart which needs action from the business's parties. These properties are used to monitor the situation of the business process. In addition, the asset comprises relationship properties to all business parties and a specific relationship property which describe the party (or parties) that should perform the next action. These properties are used to define the access rules over the asset instances and the transactions. The asset is defined in the model file (Figure 3).

- Defining the minimum number of participants in the model file utilizing enumerated type properties which are used to specify different types of the business parties.

- Considering one transaction for each Boolean properties of the asset. These transactions can change the value of the asset instance's properties whenever a progress is made in the process flow. Moreover, the model file has a specific transaction to initiate a new asset instance.

- Expanding the transaction processor functions in the script file, in a way that they not only describe the detail of the previously defined transaction, but they also imply and follow the logical sequences of the business process model.

- Describing the access rules over all asset instances and transactions based on the business model's laws. The Access Control Language can be used to define any access control method, inside the access control file (Figure 3).

In addition, we determined the feasibility of execution of the business processes on the blockchain using the suggested approach, by examining a real-world business model namely Order Processing (Figure 2). The result shows that the deployed business network on the blockchain completely ensures the correctness of the approved transactions and provides guaranteed access control over the business data in the network.

In summary, executing the collaborative business processes on the permissioned access-controller blockchain offers following benefits:

- In a permissioned blockchain, only approved individuals or legal entities can join to the network

- The blockchain offers a distributed tamper-proof and shared ledger that makes occurrence of any kind of deception almost impossible. Therefore, different parties can follow and track the situation of the process and its detail accurately.

- Using the consensus algorithm, the business parties do not need to trust any single entity or any trusted third party any more.

- The smart contracts can be used to implement the logical sequences of the business process flow and makes parties able to do appropriate possible actions. Moreover, they can be used to perform automatic transactions as well.

- Based on the requirements of the business model, any access control method such as mandatory or role-based access control can be utilized to define the access levels over parties' information, their assets, transactions, and the process flow's data.